\documentclass{emulateapj}
\newcommand{\Mpc}{\mbox{ Mpc}}

\newcommand{\yr}{\mbox{ yr}}

\newcommand{\Myr}{\mbox{ Myr}}
\newcommand{\secinv}{\mbox{ s$^{-1}$}}

\newcommand{\Lsun}{\mbox{ L$_\odot$}}

\newcommand{\lya}{Ly$\alpha$ }

\newcommand{\bq}{\begin{equation}}
\newcommand{\eq}{\end{equation}}
\newcommand{\bqa}{\begin{eqnarray}}
\newcommand{\eqa}{\end{eqnarray}}
\def\VEV#1{\left\langle #1\right\rangle} 

\begin{document}

\title{Constraints on Quasar Lifetimes and Beaming from the \ion{He}{2} \lya Forest}

\author{Steven R.  Furlanetto\altaffilmark{1} \& Adam Lidz\altaffilmark{2}}

\altaffiltext{1}{Department of Physics \& Astronomy, University of California Los Angeles; Los Angeles, CA 90095, USA; sfurlane@astro.ucla.edu}

\altaffiltext{2}{Department of Physics \& Astronomy, University of Pennsylvania; Philadelphia, PA 19104, USA}

\begin{abstract}
We show that comparisons of \ion{He}{2} \lya forest lines of sight to nearby quasar populations can strongly constrain the lifetimes and emission geometry of quasars.  By comparing the \ion{He}{2} and \ion{H}{1} \lya forests along a particular line of sight, one can trace fluctuations in the hardness of the radiation field (which are driven by fluctuations in the \ion{He}{2} ionization rate).  Because this high-energy background is highly variable -- thanks to the rarity of the bright quasars that dominate it and the relatively short attenuation lengths of these photons -- it is straightforward to associate features in the radiation field with their source quasars.  Here we quantify how finite lifetimes and beamed emission geometries affect these expectations.  Finite lifetimes induce a time delay that displaces the observed radiation peak relative to the quasar.  For beamed emission, geometry dictates that sources invisible to the observer can still create a peak in the radiation field.  We show that both these models produce substantial populations of ``bare" peaks (without an associated quasar) for reasonable parameter values (lifetimes $\sim 10^6$--$10^8 \yr$ and beaming angles $\la 90\arcdeg$).  A comparison to existing quasar surveys along two \ion{He}{2} \lya forest lines of sight rules out isotropic emission and infinite lifetime at high confidence; they can be accommodated either by moderate beaming or lifetimes $\sim 10^7$--$10^8 \yr$.  We also show that the distribution of radial displacements between peaks and their quasars can unambiguously distinguish these two models, although larger statistical samples are needed.
\end{abstract}
  

\section{Introduction} \label{intro}

Quasar spectra provide powerful probes of both the source's properties and the intergalactic medium (IGM).  One particularly useful aspect is the so-called ``proximity effect," which describes the highly-ionized zone surrounding each bright quasar.  Measuring the transition from this zone to the more uniform ionizing background characteristic of the average IGM provides an estimate of the magnitude of that background as well as source properties like the quasar luminosity, lifetime, variability, and emission geometry.

The classical proximity effect test uses the zone along the line of sight to a bright quasar by measuring the \ion{H}{1} \lya forest.  The increased radiation background causes excess transmission in the forest near to the quasar, which one can model to extract the ionization rate of \ion{H}{1}, $\Gamma_{\rm HI}$ \citep{bajtlik88, scott00}.  However, the \ion{H}{1} proximity zone typically only spans a few Mpc, within the overdense neighborhood of the quasar's massive host galaxy, and it is difficult to disentangle the intrinsic transmission bias of this region \citep{faucher08-prox}.  

Another flavor, the so-called ``transverse proximity effect," can be more powerful.  Here, we measure the impact of radiation from a foreground quasar on a different \lya forest skewer: as the line of sight passes the foreground quasar, the forest should show an enhanced ionizing background.  However, if the foreground quasar's light is beamed -- toward the observer, in this case -- or if the quasar has a short enough lifetime, its radiation may not even intersect the \lya forest skewer at all, or at least it may not strike it at the point of closest approach.  Thus the pattern of enhancements near foreground quasars should reveal information about the lifetime, beaming, and variability of quasars \citep{crotts89, moller92, adelberger04, schirber04, croft04}.\footnote{The classical proximity effect is less sensitive to these effects because all of the information comes from a single skewer (and hence a single light path).}

With the \ion{H}{1} \lya forest, one still only \emph{expects} to see an enhancement in the biased environment very close to the quasar (within $\la 3 \Mpc$) \citep{faucher08-prox, hennawi06}.  Such close pairs of quasars are rare, even in modern data sets.  Only recently have large surveys and more careful analysis led to the detection of the transverse effect: \citet{goncalves08} used higher-ionization metal line ratios to measure lifetimes $\sim 3 \times 10^7 \yr$ with no evidence for anisotropic emission, while  \citet{kirkman08} searched 130 quasar pairs, with separations $\la 3 \Mpc$, and found no increased transmission in the foreground but \emph{decreased} transmission in the background, which they explain by appealing to rapid ($\sim 10^6 \yr$) variability.

In this paper, we argue that the \ion{He}{2} \lya forest offers a better measurement of the transverse proximity effect.  It has two important advantages.  First, the \ion{He}{2} ionization rate (for which we will use $\Gamma$ as a shorthand for $\Gamma_{\rm HeII}$) is much more variable than $\Gamma_{\rm HI}$ \citep{fardal98, maselli05, bolton06, meiksin07, furl09-hefluc, furl09-heps}.  Each quasar dominates the local high-energy radiation field in a much larger region -- tens of comoving Mpc -- than it does for \ion{H}{1}, so the signature is much easier to identify.  Second, we can compare the \ion{He}{2} and \ion{H}{1} forests in order to measure the \emph{hardness} of the ionizing background.  This eliminates the dependence on density (and temperature), which affect both the \ion{He}{2} and \ion{H}{1} fractions in the same way, and removes any ambiguity due to the quasar environment \citep{faucher08-prox}.  To the extent that the \ion{H}{1}-ionizing background is uniform, we can also therefore cleanly measure the \ion{He}{2}-ionizing background.

These two factors make the detection and extraction of physical quantities much simpler with the \ion{He}{2} forest.  Indeed, the transverse proximity effect has already been detected along two different lines of sight \citep{jakobsen03, worseck06, worseck07}, even with low signal-to-noise spectra.  Here we will show that the method provides great discriminating power between quasar models with finite lifetimes and/or anisotropic emission.

This paper is organized as follows.  In Sections \ref{bare-life} and \ref{bare-beam} we use toy models of quasars with finite lifetimes and beaming to compute the fraction of peaks in the radiation field that have nearby observable quasars.  In Section \ref{angdistbn} we show that the angular distribution of the peak-quasar associations can distinguish these two sets of models.  Finally, we compare to existing observations in Section \ref{obs} and conclude in Section \ref{disc}.

All distances are quoted in proper units unless otherwise specified.

\begin{figure*}
\plottwo{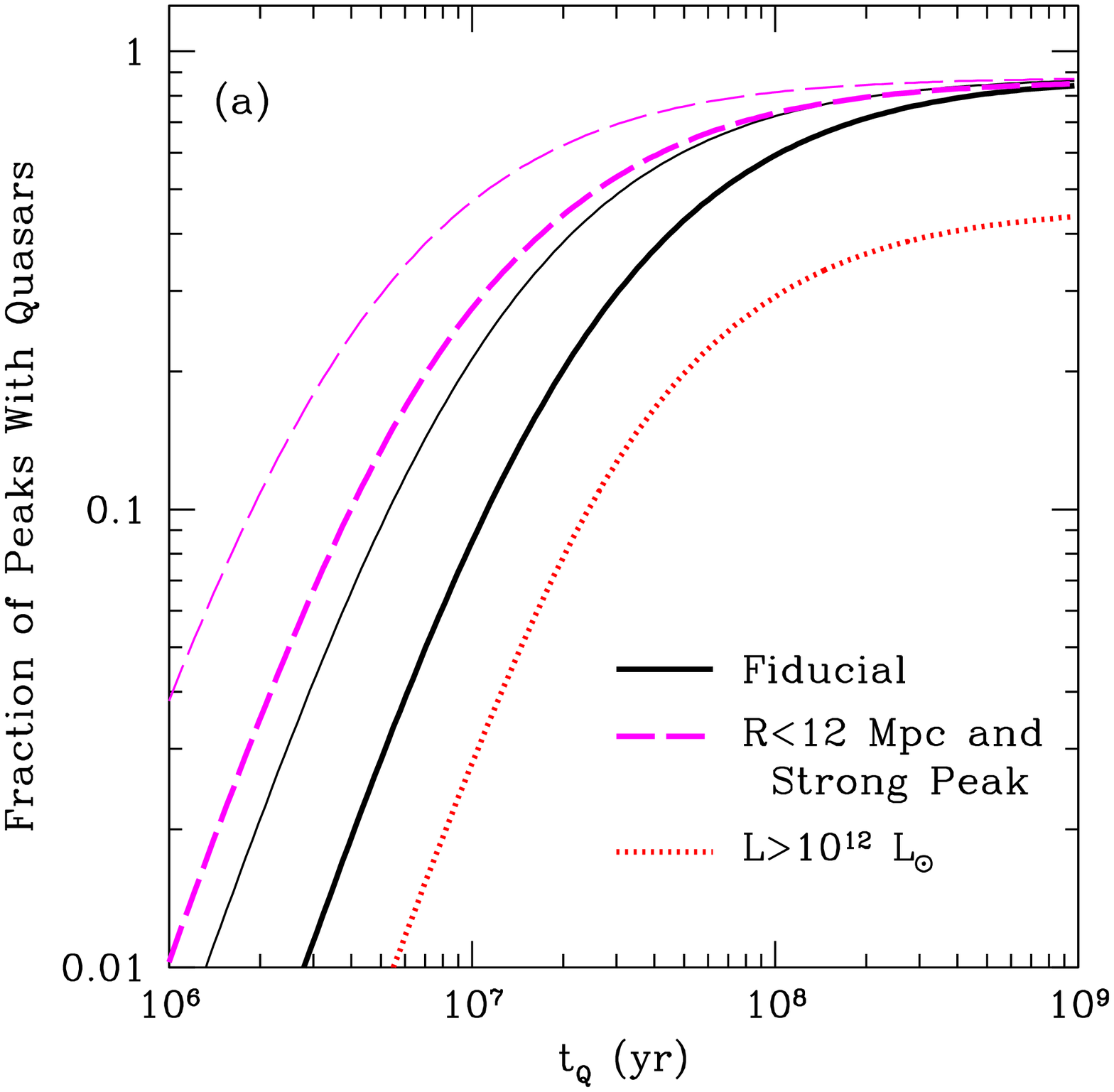}{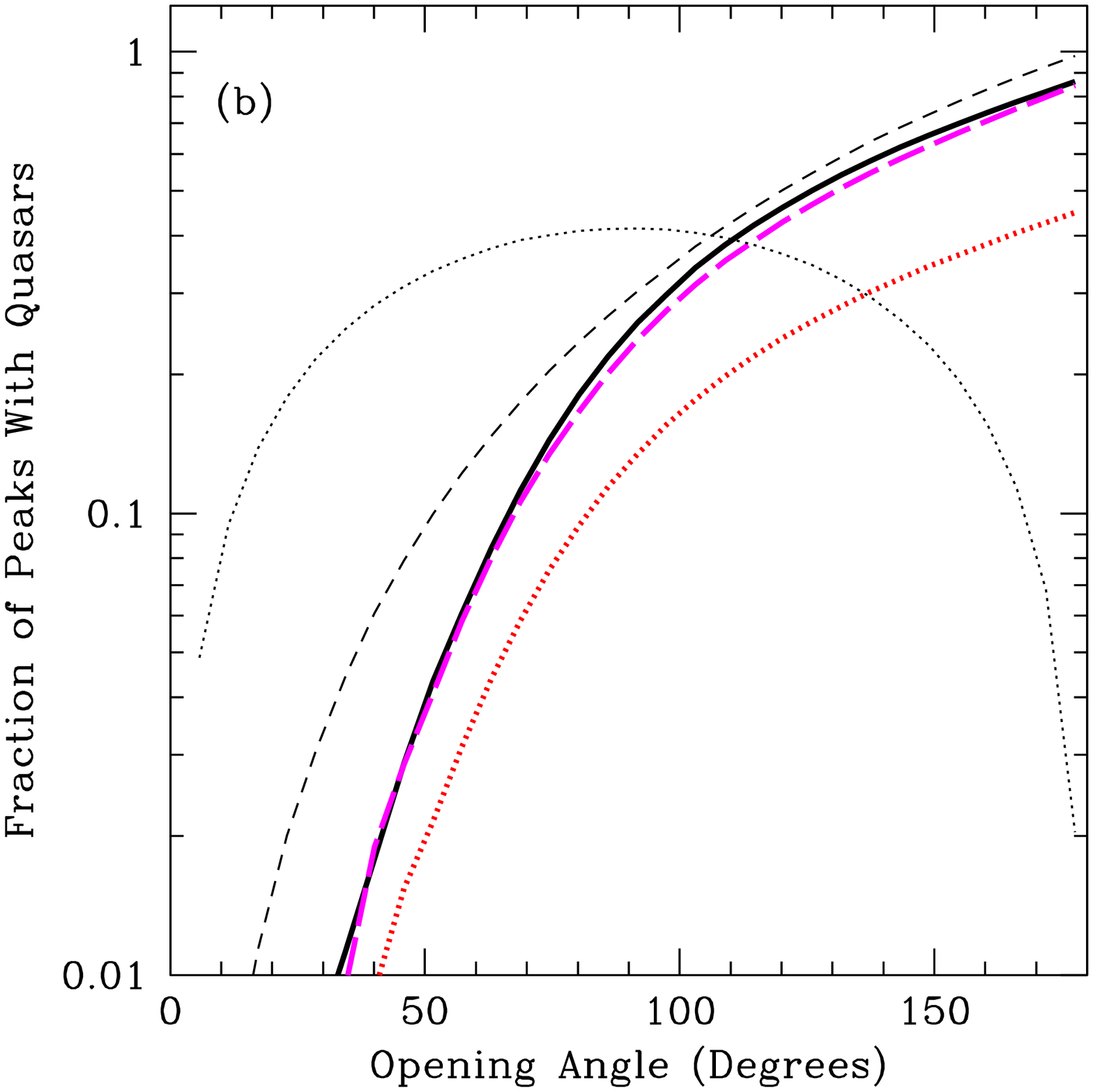}
\caption{Fraction of radiation field peaks with direct quasar associations for models with finite lifetimes (a) and beamed emission (b).  In each panel, the thick solid curve shows our fiducial model (see text), the thick dotted curve takes a higher survey luminosity threshold, and the thick dashed curve shows the predictions for a survey spanning only 12 comoving Mpc around the \lya forest skewer and in which only peaks with $\sigma > 8$ are identified.  In (a), the (upper) thin curves take $r_0=200$ comoving Mpc.  (This parameter does not affect the results with beaming.)  In (b), the thin curves show the absolute fractions of all quasars that are visible to the observer (short-dashed) and that are not visible but produce measurable peaks in the forest (dotted).}
\label{fig:barepeak}
\end{figure*}

\section{Quasars and Radiation Peaks:  Finite Lifetimes}
\label{bare-life}

Here we will use toy models and simple statistics to illustrate the power of the \ion{He}{2} transverse proximity effect.  Our basic approach is to measure the fraction of peaks in the radiation field identified along a \lya forest line of sight with \emph{observable} quasars sufficiently nearby to be clearly identified as the peak's source.  For isotropically emitting quasars with infinite lifetimes, \emph{all} peaks must have such an association, but some will disappear if quasars are beamed (because many of the sources will not be oriented toward the observer) or if they have finite lifetimes (because many of the sources will have shut off already).

For the purposes of this paper, we take a very simple toy model in which only a single (nearby) quasar induces fluctuations in the \ion{He}{2}-ionizing background; we assume that the cumulative background from other sources is uniform.  This is certainly a simplification \citep{furl09-hefluc}, but we will defer a more comprehensive model to future, more detailed work.  Figure~1 of \citet{furl10-taufluc} shows that this is a reasonable approximation in most cases of interest (i.e., near strong peaks in the radiation field).

We define a ``peak" in the radiation field as any point at which the \ion{He}{2}-ionizing background intensity $J > \sigma \VEV{J}$, where $\sigma$ is a constant.  We will usually choose $\sigma=1$, because the \emph{median} ionizing background is a factor $\sim 3$ less than the mean \citep{furl09-hefluc}.  

In practice, such peaks will be identified as minima in the ``hardness ratio" $\eta = N_{\rm HeII}/N_{\rm HI} \propto \Gamma_{\rm HI}/\Gamma$, the ratios of column densities in singly-ionized helium and neutral hydrogen within a given absorber \citep{miralda93}.  We implicitly assume that $\Gamma_{\rm HI}$ is uniform, so $\eta \propto 1/\Gamma \propto 1/J$. This appears to be an excellent approximation at $z \la 4$ \citep{meiksin03, croft04}.  This approach has the added benefit of eliminating any uncertainty in the density or temperature structure of the forest. Note, however, that regions so close to their sources so as to be within both the proximity zones of \ion{He}{2} and \ion{H}{1} may be missed by this technique.

We then define $r_p$ as the maximum transverse distance within which a quasar can sit and still produce a radiation peak. If the quasars are Poisson-distributed and the mean free path of an ionizing photon is $r_0$, the average radiation field is $\VEV{J} = 3 \bar{N}_0 J_\star$ \citep{meiksin04}, where $\bar{N}_0$ is the average number of quasars inside one attenuation zone (with radius $r_0$), $J_\star = \VEV{L}/(4 \pi r_0)^2$, and $\VEV{L}$ is the mean quasar luminosity (averaged over the luminosity function, which we take from \citealt{hopkins07}).  

There is unfortunately considerable disagreement on the mean free path of these photons. At $z \sim 2.5$, models predict that $r_0 \sim 45$--$200$ comoving Mpc (e.g., \citealt{bolton06, faucher09}).  One method uses the IGM density distribution in simulations and calibrates to the observed average optical depth \citep{bolton06}, and it reproduces the abundance of \ion{H}{1} Lyman-limit systems from \citet{storrie94}.  However, it relies on ad hoc geometric assumptions about the absorbers.  \citet{fardal98} and \citet{faucher09} use photoionization modeling of absorbers in the \ion{H}{1} forest; the latter reproduces the more recent mean free path measurements (for photons that ionize \ion{H}{1}) of \citet{prochaska09} and \citet{songaila10}.  However, this method relies sensitively on the \ion{H}{1} forest absorbers with $N_{\rm HI} \sim 10^{15}$--$10^{17} \mbox{ cm}^{-2}$, which are very difficult to measure, and other estimates give much smaller values \citep{fardal98}.  The more recent data is calibrated to $z \sim 3.7$, so it also cannot account for possible evolution to the redshifts of interest, $z \sim 2.5$.  Neither method takes into account the substantial fluctuations of the \ion{He}{2} ionizing background \citep{furl09-hefluc}.  

Other arguments suggest that $r_0$ lies in the middle of this range.  First, let us assume that the abundance of \lya forest absorbers follows a power law $\propto N_{\rm HI}^{-3/2}$, close to observational limits (e.g., \citealt{fardal98}).  In the optically thin limit with uniform radiation backgrounds, the hardness ratio $\eta$ is spatially constant.  We can therefore use the distribution of \ion{H}{1} absorbers to relate the mean free paths of \ion{H}{1} and \ion{He}{2} ionizing photons ($r_{\rm HI}$ and $r_0$ respectively):
\bq
r_{\rm HI} = r_0 \sqrt{\eta/4}.
\eq
\citet{faucher09} estimate that $r_{\rm HI} = 85 ([1+z]/4)^{-4} \Mpc$, which matches the \citet{prochaska09} measurement at $z=3.6$ very well, although the redshift evolution is uncertain.  In that case, $\eta \sim 40$--$80$ \citep{shull04, zheng04, fechner06, fechner07} implies $r_0 \sim 110$--$160 \Mpc$ at $z \sim 2.5$.  

A separate argument comes from matching the (measured) emissivity of the quasar population with the observed optical depth of the \ion{He}{2} forest.  In the fluctuating Gunn-Peterson approximation, $r_0 \ga 100 \Mpc$ overproduces the ionization rate compared to observations \citep{dixon09}, although there is an uncertain correction factor in this model.

We therefore take $r_0=45$ comoving Mpc as a fiducial model but show results for larger values as well.\footnote{An alternative approach is to scale the proximity zone to the \ion{He}{2} ionization rate, $\Gamma$.  In that case, we remove the uncertainty in $r_0$ but replace it with equivalent uncertainty in $\Gamma$.}   Then we find $r_p \approx r_0 \sqrt{L / \tilde{L}}$, where 
\bq
\tilde{L} = 3 \bar{N}_0 \sigma \VEV{L} \approx 5 \times 10^{12} \sigma \left[ {r_0 (1+z) \over 45 \Mpc} \right]^3 \Lsun.
\eq
is a characteristic luminosity.  (Here the factor $1+z$ converts our physical coordinates into comoving units.)  Quasars must sit inside this region \emph{and} intersect the \lya forest skewer in order to produce an observable peak.  For reference, a bright quasar ($\sim 10^{12} \Lsun$) has $r_p \sim r_0/\sqrt{5} \sim 20$ comoving Mpc with our fiducial value for $r_0$; $r_p \sim 12$ comoving Mpc for $r_0 = 150$ comoving Mpc.

If quasars have a finite lifetime $t_Q$, the induced radiation peak along a nearby line of sight can be displaced radially from the quasar location, and -- because of light travel time delay -- the quasar also may not be visible when the peak is observed. To compute the probability to have a ``bare" peak (i.e., without an associated quasar), we note that the quasar and peak are \emph{both} visible 1) once the peak has moved within a distance $r<r_p$ of the source and  2) before $t_Q$ has elapsed.  The first corresponds to a minimum time after the quasar appears of
\bq
t_{\rm min} = { r_p - \sqrt{r_p^2 - r_\perp^2} \over c},
\label{eq:tmin}
\eq
where $r_\perp$ is the quasar's impact parameter from the \lya forest skewer.  Before $t_{\rm min}$, the peak along the \lya forest skewer is at least $r_p$ in front of the quasar and is invisible.  Of course, if $t_{\rm min} > t_Q$, the peak never reaches the proximity zone during the quasar's lifetime, so there cannot be an association.

After the quasar shuts off, the peak will continue to drift farther from the observer as the corresponding light travel time delay increases.  The peak will reach the back of the proximity zone after a time delay, relative to the last quasar photon seen by the observer, of
\bq
\Delta t_{\rm max} = { r_p + \sqrt{r_p^2 - r_\perp^2} \over c}.
\eq

If all quasars have a fixed luminosity, so that $r_p$ is constant across the population, then the fraction of peaks with quasar associations is $F_{\rm lt} = N_{pq}/(N_{pb} + N_{pq})$, where
\bqa
N_{pq} & = & \pi \int_0^{r_{\rm max}} dr_\perp r_\perp c [t_Q - t_{\rm min}(r_\perp)] \label{eq:Npq} \qquad \mbox{and}\\
N_{pb} & = & \pi \int_0^{r_p} dr_\perp r_\perp c [\Delta t_{\rm max}(r_\perp) - {\rm max}(t_{\rm min} - t_Q,\,0)]
\label{eq:Npb}
\eqa
are respectively the volumes within which quasars can sit and have a visible peak (or not). Here $r_{\rm max}$ enforces the requirement that $t_{\rm min} \le t_Q$; it is
\bq
{r_{\rm max} \over r_p} = \sqrt{ 1 - \left( 1 - \tilde{r}_Q \right)^2 }
\eq
if $\tilde{r}_Q \equiv c t_Q/r_p < 1$ and unity otherwise.  The final factor in $N_{pb}$ accounts for the time lag between the quasar shutting off and the peak reaching the proximity zone, in cases where $t_Q < t_{\rm min}$.  

The integrals in equations~(\ref{eq:Npq})-(\ref{eq:Npb}) can be performed analytically; the results scale roughly as $r_p^2$ from the cylindrical geometry of the problem, with a suppression at large luminosity because of the finite lifetime limits (even though bright quasars have large proximity zones, they can still shut off before the peak becomes visible).  Of course, we must actually integrate $N_{pb}$ and $N_{pq}$ over the quasar luminosity function.  Because these factors scale like $r_p^2 \propto L$, the most luminous quasars are by far the most important for generating peaks in the radiation field.

Figure~\ref{fig:barepeak}\emph{a} shows the resulting fractions for several different mock surveys.  In order to better mimic real surveys, we take a minimum luminosity threshold when calculating the fraction of observed associations, though not when calculating the total number of peaks (i.e. we limit $L$ in the numerator of $F_{\rm lt}$ but not in the denominator).  We do not (yet) limit the spatial extent of the survey, instead assuming that all quasars can be identified out to the appropriate $r_p(L)$, which is in principle measurable from each quasar's luminosity.  In practice, this would require a survey that is deepest near the \lya forest line of sight.

The thick solid curve takes $L > 10^{11} \Lsun$ and $r_0=45$ comoving Mpc at $z=2.5$.  The fraction of associations increases rapidly at $t_Q \la 3 \times 10^7 \yr$ and then flattens out at larger lifetimes.  For small lifetimes, we expect the result to be $\sim \tilde{r}_Q^2$, which is the fraction of the proximity zone for which the light travel time is less than $t_Q$.  This is indeed roughly correct; once $\tilde{r}_Q \ga 1$, the curves flatten significantly because the more distant quasars only provide observable peaks for brief windows of time anyway.

The dotted curve shows how the fraction varies with the survey depth, taking $L>10^{12} \Lsun$.  Clearly a shallower survey strongly reduces the number of observed associations.  However, note that decreasing the limit below $10^{11} \Lsun$ has very little effect.  One need only identify those quasars responsible for strong peaks, which are primarily bright and moderate luminosity sources.  To produce a peak, faint sources must already be so close to the line of sight that their available volume is small.  However, although fewer peaks have associations,  the \emph{variation} of the curves with $t_Q$ is relatively constant with $L$, so a wide, shallow survey may be just as effective as a deeper one, if one is confident enough about modeling the fainter quasar population.

The thin solid curve takes $r_0=200$ comoving Mpc; we find that the fraction of associations is roughly proportional to $r_p^2 \propto r_0^{-1}$ at short lifetimes.  A larger attenuation length increases the fraction of peaks with quasars, because the additional sources illuminating each point effectively decrease $r_p$ (and hence time delay effects) in order to overcome the background from the other sources.  The dependence is substantial, so a more accurate estimate of the mean free path will be essential for detailed constraints.

Finally, the long-dashed curves assume a survey comparable to \citet{worseck07}; again the thick and thin curves take $r_0=45$ and 200 comoving Mpc, respectively.  We take $L>10^{11} \Lsun$, only include peaks with $\sigma > 8$, and only identify quasars within 12 comoving Mpc of the \lya forest skewer.  These factors decrease the dependence on $t_Q$ (and hence make the survey less sensitive), largely because of the high peak threshold:  quasars must be very close to the line of sight in order to produce such strong peaks.  These nearby sources do not provide much constraining power, because the light travel time is then small compared to $t_Q$; surveys for bright quasars at distances near $r_p$ are most efficient.  Nevertheless, short lifetimes ($t_Q \sim 10^6 \yr$) would imply very few observable associations with this kind of survey (see Section \ref{obs}), and very long lifetimes $t_Q \ga 10^8 \yr$) would imply almost perfect association.

\section{Quasars and Radiation Peaks: Beaming} \label{bare-beam}

We now switch focus to models with infinite lifetime but anisotropic quasar emission (or ``beaming").  For concreteness, we will use a simple biconical emission model, in which two oppositely-directed beams each have opening angle $\Omega$.  

In this case the calculation is conceptually simple:  how often does a beam that remains invisible to the observer intersect the \lya forest line of sight between the front and back edges of the proximity zone?  Some visible quasars will have such intersections, some will have none, and some invisible quasars will still cause peaks; our accounting must include all these possibilities.  (Note that quasars whose beams do not intersect \emph{either} the observer's line of sight or the \lya forest skewer remain entirely invisible and can be ignored.)

We use a Monte Carlo model to compute these probabilities.\footnote{Without the limits imposed by the proximity zone, an analytic calculation is straightforward -- in fact \emph{every} visible quasar produces a peak somewhere -- but requiring some part of the beam to strike the skewer within $r_p$ of the quasar's location makes such a model unwieldy.}  The thin curves in Figure~\ref{fig:barepeak}b show (1) the fraction of quasars that are visible to the observer and produce visible peaks (short-dashed curve) and (2) the fraction that are invisible to the observer but still produce a peak (dotted curve).  The former simply increases with $\Omega$, of course.\footnote{Note again that all quasars produce ``peaks" somewhere along the line of sight, but we do not count them unless they are within the proximity zone.}  The latter initially increases (as quasars become more likely to intersect the skewer) and then decreases (as quasars become more likely to be seen by the observer).

To generate predictions relevant to observations, we must include quasars of all luminosities, as before.  However, if we assume (as in our fiducial model) that we detect all quasars within their respective $r_p(L)$, then the results become independent of luminosity because each quasar is treated identically (this differs from the finite lifetime case, where $\tilde{r}_Q$ introduces a second physical scale).  If, however, $r_{\rm max}$ is fixed (as in a real survey), then the integration over the luminosity function becomes necessary.  Similarly, if the survey is not infinitely deep, the minimum luminosity threshold introduces luminosity dependence because some associations with faint quasars will be missed.

The thick solid curve in Figure~\ref{fig:barepeak}b shows results for a survey with $r_{\rm max} = r_p(L)$ and $L > 10^{11} \Lsun$.  The number of bare peaks can be substantial; roughly two out of three peaks will have no visible association if $\Omega \sim \pi/2$, with the ratio declining very rapidly at smaller opening angles.  As in the finite lifetime case, a shallower survey misses associations.  However, in this case the shape of the curve as a function of $\Omega$ does change with the luminosity threshold, becoming slightly more sensitive to variations in $\Omega$ as the survey deepens.

\begin{figure*}
\plottwo{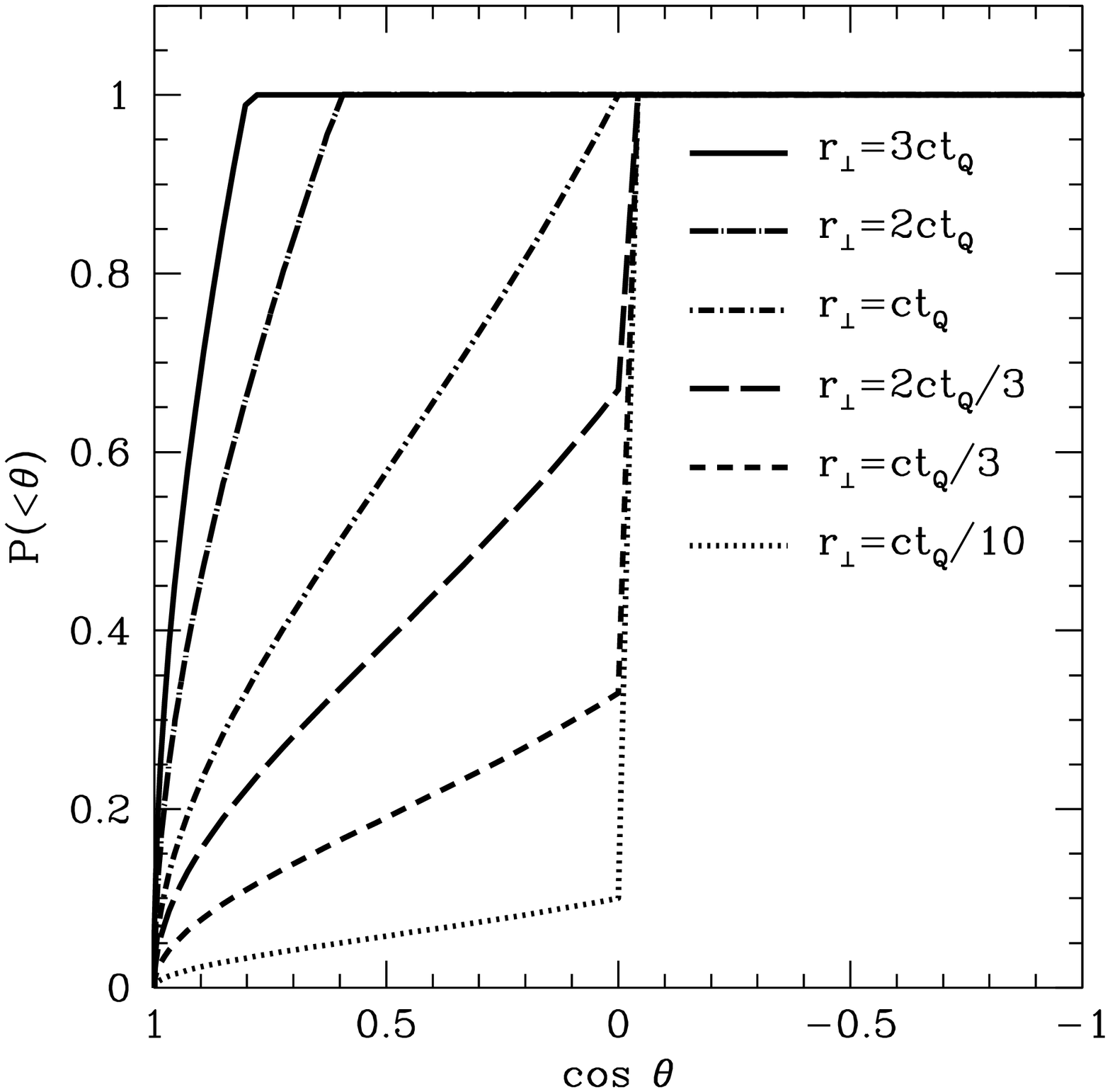}{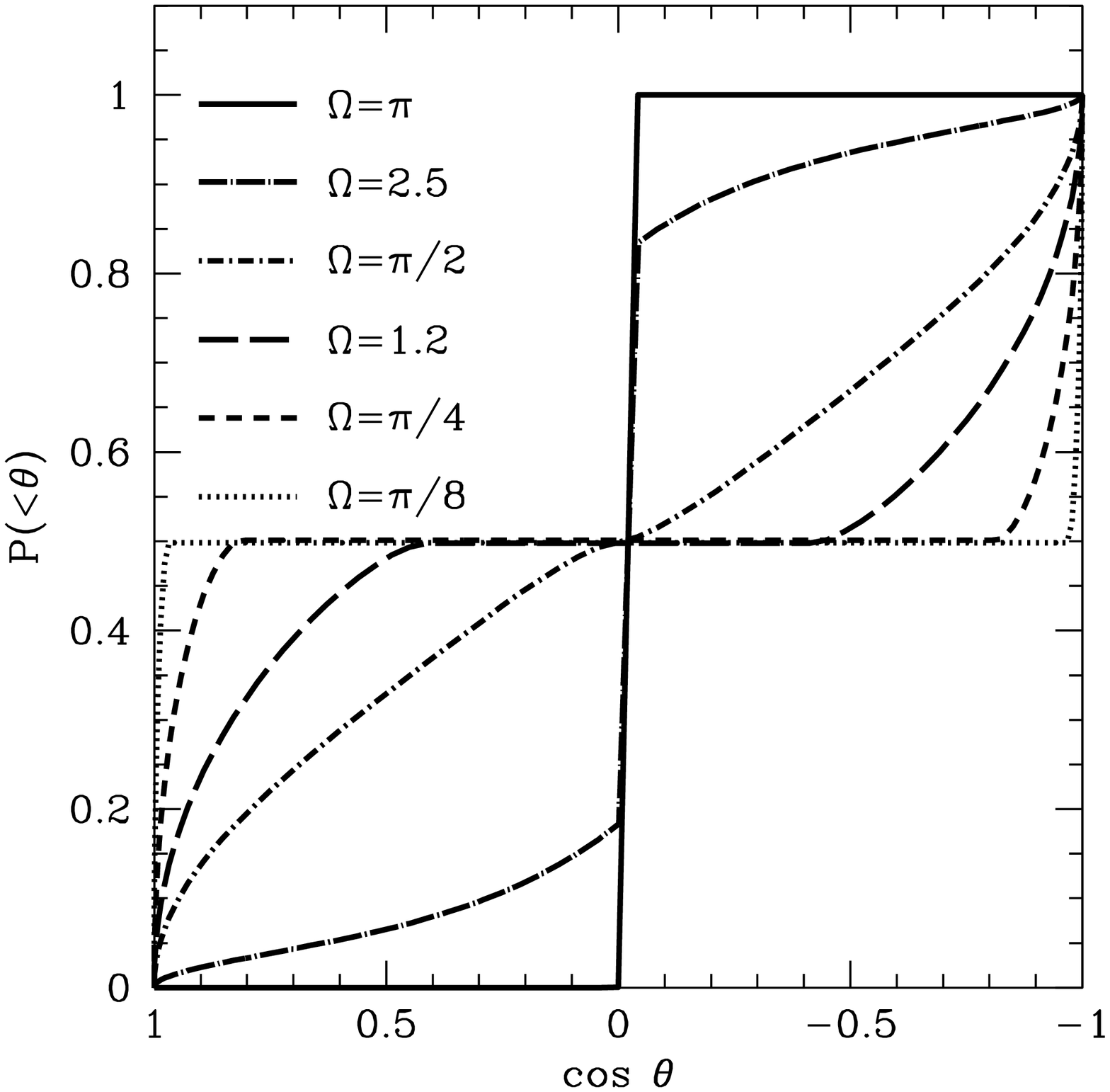}
\caption{Cumulative distribution function of the peak angle $\theta$ for quasars with finite lifetimes (\emph{left}) and biconical beaming (\emph{right}).   In the beaming case, the distributions are independent of distance from the quasar.  In the finite lifetime case, we show several different impact parameters between the \lya forest skewer and the quasar, scaled to the light travel time over the quasar's lifetime, $ct_Q$.}
\label{fig:qsopeak}
\end{figure*}

The long-dashed curve again shows a survey similar to \citet{worseck07}, which only includes strong peaks and nearby quasars.  Because there is no physical scale other than $r_p$ in the problem, simply increasing $\sigma$ -- which affects all quasars equally -- does not affect the fraction of bare peaks.  On the other hand, limiting $r_{\rm max}$ does, because it eliminates the possibility of finding more distant luminous counterparts.  However, with the large $\sigma$ imposed here nearly all peaks are sourced by quasars within the surveyed region, so the finite area makes only a small difference to the final curve, which is mostly determined by the depth of the survey.

It is worth emphasizing that the beaming case is somewhat more robust to observational uncertainties, because it does not depend on $r_0$ (at least in our simplified model).

\section{Peak-Quasar Associations} \label{angdistbn}

So far we have shown that the relative number counts of ``bare" peaks and those with associated quasars depends on the lifetime and beaming angle. Here we briefly show that the relative radial locations of the peaks and their associated quasars can provide additional evidence and in particular distinguish these two scenarios.

Clearly, if each quasar emits isotropically over an infinite lifetime, it illuminates every point in the universe, so the brightest point along a nearby \lya forest skewer (at a radial distance from the observer $r_{\rm peak}$) will lie at the same radial distance as the quasar:  i.e., if $\theta$ is the angle between the \lya forest skewer and the ray joining $r_{\rm peak}$ and the quasar, then $\theta=\pi/2$ for every source, barring complex radiative transfer effects and errors in localizing the quasar and peak.  Deviations from this simple expectation therefore indicate more complex quasar properties, such as finite lifetimes or beaming.

\subsection{A Finite Lifetime} \label{ideal-lifetime}

With a finite lifetime, a quasar visible through both its direct emission and through its influence on a point in the IGM must satisfy a time delay criterion:  the difference in light travel time along these two paths $\Delta t$ must be no greater than $t_Q$.  In terms of the angle $\theta$, this time delay is
\bq
\Delta t = (r - d_{\rm LOS})/c = (1 - \cos \theta)r/c = \left( {1 - \cos \theta \over \sin \theta } \right) {r_\perp \over c},
\label{eq:Deltat}
\eq 
where $r$ is the total distance from the quasar to the nearest point that it illuminates and $d_{\rm LOS}$ is the radial distance between that same point and the quasar.  This is an \emph{increasing} function of $\theta$, so points closest to the observer have the least delay,\footnote{We assume here that the quasar is effectively at an infinite distance from the observer.} and it increases monotonically as $\theta$ approaches $\pi/2$.  

We will construct the probability distribution of the angle $\theta$ in a finite lifetime model (recall that it is a delta function at $\theta=\pi/2$ for the fiducial infinite  lifetime model).  If the quasar turned on a time $t$ in the past, only points with $\Delta t < t$ are illuminated, and the peak is at the point with the largest $\theta$.  Assuming a uniform distribution of quasar ages, the cumulative distribution of peak locations is therefore
\bq
P(<\theta) = {\Delta t \over t_Q} = \left( {1 - \cos \theta \over \sin \theta } \right) {r_\perp \over c t_Q}.
\eq

The left panel of Figure~\ref{fig:qsopeak} shows this distribution for several impact parameters $r_\perp$, scaled to $c t_Q \approx 30 (t_Q/10^8 \yr) \Mpc$.  Two points are immediately obvious.  First, $\theta < \pi/2$ always; once the quasar illuminates this point of absolute closest approach, the peak must remain there until the quasar shuts off.  Thus we expect a clear asymmetry between the forward and backward directions.  Second, the distribution depends strongly on the impact parameter from the skewer.  Nearby quasars have short delays, so $\theta=\pi/2$ is most common.  But those at relatively large impact parameters only rarely have the timing just right to attain $\theta=\pi/2$.  Thus faint quasars -- which must be nearby to influence the radiation field -- provide little additional information on finite lifetimes, and surveys for bright quasars over wide areas are most productive.

Note that here we have not restricted ourselves to quasars with visible peaks within a distance $r_p$ from the source.  The importance of this restriction depends on the impact parameter $r_\perp$; those peaks with $\cos \theta \sim 1$ will be difficult to identify in practice.

\subsection{Quasar Beaming} \label{ideal-beaming}

Now we consider the distribution of $\theta$ in a beaming model with infinite lifetime; specifically, biconical emission.  More complicated emission geometries are of course possible and can dramatically change the estimates in this section.

For a quantitative picture, we again turn to the distribution of $\theta$.  In the beamed case, quasar-peak pairs visible to the observer will typically \emph{not} have $\theta=\pi/2$, because the beams can only subtend (at most) an angle $\Omega$ from $\theta=0$ or $\pi$ (along the radial direction).  The right panel of Figure~\ref{fig:qsopeak} shows the resulting cumulative probability distributions of $\theta$ for a range of opening angles (generated with a Monte Carlo model).\footnote{Note that we normalize these distributions to unity, so we include only those sources that are both visible \emph{and} create peaks along the line of sight; the other configurations discussed in Section \ref{bare-beam} are not relevant for this test.}

As expected, when each beam has $\Omega=\pi$ (i.e., isotropic emission), the peak is always located perpendicular to the line of sight.  As $\Omega$ decreases, the probability of this configuration decreases rapidly, and by $\Omega=\pi/2$, it is vanishingly rare.  Instead, the peak drifts farther and farther from the perpendicular, because the illuminated region is oriented more and more directly toward (or away from) the observer.  

There are two clear differences from the finite lifetime case.  First, the beaming results are distance-independent, because there is no physical scale in the problem (although ``peaks" beyond $r_p$ will no longer be visible in practice).  Second, the distributions are symmetric about $\theta = \pi/2$, because we assume two beams directed in opposite directions; $\theta > \pi/2$ indicates that different beams illuminate the skewer and observer.  The second difference therefore depends on the particular model of quasars, but the first will unambiguously distinguish finite lifetimes from geometric effects.

\section{Comparison to Existing Observations} \label{obs}

So far, searches for neighboring quasars have been conducted along two of the five lines of sight with \ion{He}{2} \lya forest data.  The best sample so far surrounds the line of sight to HE 2347--4342, whose \ion{He}{2} forest has been extensively studied \citep{zheng04, shull04, fechner07, shull10}.  \citet{worseck07} searched for nearby quasars with $L_{\rm min} \ga 1$--$2 \times 10^{11} \Lsun$ and a maximum impact parameter $\sim 12$ comoving Mpc.  They discovered two nearby quasars, both with associated peaks in the high-energy radiation background (actually identified as troughs in the hardness ration $\eta$).\footnote{They also report a third quasar just below their survey limit, which we do not include to ensure completeness.  However, because it lies very close to one of the other two quasars, it actually adds no additional information.}  Three other regions of the spectrum have a radiation field at least as hard as these (with $\sigma=8$), so the fraction of associations is 2/5.  

The second skewer is toward Q0302--003.  \citet{heap00} identified a strong transmission feature near $z = 3.05$.  \citet{jakobsen03} subsequently identified a quasar at a projected distance of just $1.77$ comoving Mpc from this feature.  \citet{worseck06} later performed a deeper search (to comparable magnitude and volume limits as above) and identified one more quasar within $\sim 5 \Mpc$ (projected) of the \ion{He}{2} \lya forest line of sight and a third inside the classical proximity zone of Q0302--003 (we ignore the last, since its effects are difficult to disentangle).  

\citet{worseck06} find peaks in the hardness of the radiation background near the locations of all of these quasars.  They do not quantify the number of peaks without quasars, but visual inspection of their Fig.~7 show that the sample also contains two other hardness peaks of comparable amplitude to those with known quasars. If these are included (so that the fraction of associations is 4/9 across both lines of sight), we can estimate the likelihood of this result given the various parameter sets.  Assuming that each peak provides an independent test, and taking the probability of successfully finding a nearby quasar from Figure~\ref{fig:barepeak} (the long-dashed curves match the survey parameters reasonably well), then the probabilities to find exactly four associations in the sample are $(0.17\%,\, 2.0\%,\, 14\%,\,23\%,\, 5.1\%)$ for $t_Q = (0.3,\,0.5,\,1,\, 3,\, 10) \times 10^7 \yr$, respectively, if we take $r_0 =45$ comoving Mpc.  Thus, with this single line of sight we can rule out at high confidence lifetimes of $t_Q \la 5 \times 10^6 \yr $ (or even significant variability on those timescales) or $t_Q \ga 10^8 \yr$.

However, taking a larger mean free path ($r_0=200$ comoving Mpc) changes the constraints.  In that case, the probabilities to find exactly four associations in the sample are $(0.022\%,\,4.9\%,\, 26\%,\,8.2\%,\, 1.2\%)$ for $t_Q = (0.1,\,0.3,\,1,\, 3,\, 10) \times 10^7 \yr$, respectively.  We can still therefore rule out long lifetimes $t_Q \ga 10^8 \yr$ by the lack of perfect associations, but we cannot place as strong constraints at the short end.  Note that the constraints do not improve much more at large $t_Q$, because the distribution flattens out when $\tilde{r}_Q$ exceeds unity for bright quasars.

We also find probabilities of $(0.11\%,\,8.5\%,\, 9.5\%,\, 0.5\%)$ to find exactly four associations if $\Omega = (60\arcdeg,\, 90\arcdeg,\, 135\arcdeg, 180\arcdeg)$, respectively.\footnote{The small chance for isotropic emission ($\Omega=180\arcdeg$) accounts for quasars below the minimum luminosity limit.}  Thus here we can place strong constraints, ruling out any beaming scenarios with $\Omega \la 90\arcdeg$ as well as isotropic emission (at least with the infinite lifetime approximation); these are independent of the uncertainty in $r_0$ in our simple model where the nearest source provides a clear peak.  Note that finite lifetimes and beaming both decrease the observed ratios, so combined models constrain small $\Omega$ even more tightly (although isotropic emission with finite lifetimes is permitted).

In this analysis we have used the low signal-to-noise \emph{Far Ultraviolet Spectroscopic Explorer} (\emph{FUSE}) \ion{He}{2} spectra available at the time of the quasar searches \citep{worseck06, worseck07} in order to allow a uniform sample across both spectra.  Very recently, \citet{shull10} observed HE2347--4342 with the newly-installed Cosmic Origins Spectrograph (COS) on the \emph{Hubble Space Telescope}.  The remarkable improvement in the \ion{He}{2} \lya forest spectrum clearly shows the peaks and troughs in the hardness ratio (their Fig.~7), and in principle this makes our test much easier.  In this case it is very easy to identify peaks with $\sigma \ga 5$; in fact the spectrum is so good that small-scale radiative transfer effects likely become important, since multiple peaks are often visible very close together.  It is therefore somewhat difficult to pick out the \emph{independent} peaks in the spectrum, so a more detailed simulation is likely necessary in order to use this data most efficiently.

For a simple estimate, we count the number of regions of extent $r_p \sim 20$ comoving Mpc with peaks inside them and use that as a proxy for the true number of independent peaks.  We count seven such regions in the range $z=2.4$--$2.72$ (although several are very near our threshold); the spectrum cuts off below this redshift, and at higher values \ion{He}{2} reionization may interfere with our simple model \citep{dixon09, furl10-taufluc, shull10}.  Only one of the \citet{worseck07} associations sits in this range, so along this segment we have 1/7 peak-quasar associations.  Using $\sigma=5$, we find the probability of this result to be $(4.4\%,\,24\%,\, 35\%,\,8.0\%,\, 0.43\%)$ for $t_Q = (0.1,\,0.3,\,1,\, 3,\, 10) \times 10^7 \yr$, respectively, and $(39\%,\,27\%,\, 3.2\%,\, \approx 0)$ or if $\Omega = (60\arcdeg,\, 90\arcdeg,\, 135\arcdeg, 180\arcdeg)$, respectively.  This spectrum is more tolerant of short lifetimes and tight beaming but less tolerant of long lifetimes and isotropic emission.  As emphasized above, however, the data here are sufficiently good that simulations are clearly necessary to understand it.  The difficulty of classifying peaks and sources in this new data points to a statistical approach as most robust.

For most of these quasars, the local minima in $\eta$ coincide in redshift with the quasars themselves, which implies that isotropic emission and relatively long lifetimes are good approximations.  However, one source (at $z=2.69$ toward HE 2347--4342) has a peak somewhat in front of it according to the \emph{FUSE} spectrum.  \citet{worseck07} estimate that $t_Q \ga 25 \Myr$ from this coincidence.  According to our models, the position in front of the quasar is suggestive of a finite lifetime as an explanation, but it could be explained equally well by a beaming model.  On the other hand, the COS spectrum shows a peak at $z=2.69$ as well as the one at lower redshifts, which would easily be consistent with longer lifetimes or isotropic emission.

In any case, confusion from more distant sources as well as redshift errors can easily mimic this level of displacement in individual sources, so statistical samples are necessary for strong constraints based on the relative positioning of peaks and quasars.

\section{Discussion} \label{disc}

We have examined how the transverse proximity effect, observed through a combination of the \ion{He}{2} and \ion{H}{1} \lya forests, can help to constrain quasar properties.  Using simple toy models, we showed how finite source lifetimes and beamed emission affect the statistical association between peaks in the \ion{He}{2} ionizing background and quasars. Both scenarios can substantially decrease the probability that the source causing a given peak is visible to a distant observer.  Finite lifetimes break such associations when $ct_Q$ is less than the ``proximity radius" within which a quasar's radiation is more important than the accumulated background.  Quasars with beamed emission may or may not be visible to both the \lya forest skewer and the observer for purely geometric reasons.  As such, these models are less sensitive to the impact parameter of the quasar.

We have shown that, even with the existing data (relatively small surveys around two lines of sight), this technique suggests that $t_Q \la 10^8 \yr$ (with possibly some constraints $t_Q \ga 3 \times 10^6 \yr$ if the older sample is more appropriate for our toy model).  Similarly, it appears to rule out either very small ($\la 60\arcdeg$) opening angles and isotropic emission (at least if an infinite lifetime is assumed).  Of course, our simple toy models, which focus only on the single nearest source, are not sufficient to claim rigorous constraints.  Improved numerical or Monte Carlo models can better test the importance of the accumulated background of distant quasars, errors in quasar locations, and errors in peak detection in the \ion{He}{2} forest.  The impressive COS \ion{He}{2} spectrum from \citet{shull10} illustrates some of these difficulties, as one can clearly trace the peaks and troughs of the ionizing background, and it is difficult to determine which narrow peaks to associate with each source. Nevertheless, the constraints suggested by our toy model show that this is a very promising technique.

The mere presence or absence of a nearby quasar is also only the first bit of information from such surveys; the distribution of peak locations around quasars contains much more.  Causality dictates that sources with finite lifetimes create peaks in front of (or at worst coincident with) the source, while beamed quasars can have their peaks behind the sources.  The existing surveys have three peaks nearly coincident with their sources and one (possibly) leading; this provides weak evidence for finite lifetimes and/or anisotropy, but much better modeling is needed to interpret the data fully.

In designing a survey, we have found that the brightest neighbors are the most useful, because they have the largest proximity zones and produce the most obvious peaks. Faint quasars must be extremely close to the line of sight in order to create a peak; if they are so close, then the light travel time is small so they do not efficiently constrain finite lifetimes.  (The same is true of bright quasars very near to the line of sight; the strongest constraints on lifetimes will be provided by quasars at moderate distances from the \lya forest skewers.)

We have argued that this method has two advantages over more traditional searches involving just the \ion{H}{1} forest \citep{worseck06}.  First, the transverse proximity zone is much larger in \ion{He}{2}, because only (rare) quasars contribute to the high-energy ionizing background and because the IGM attenuation length is several times smaller.  Second, and more important, comparing the \ion{He}{2} and \ion{H}{1} \lya forests provides a direct measurement of the hardness of the radiation field, robust to variations in the underlying IGM density and temperature.  This avoids a substantial bias in the \ion{H}{1} proximity effect \citep{faucher08-prox}.

Our tentative results from these toy models are consistent with the detection of the transverse proximity effect through metal lines by \citet{goncalves08}, who estimated $t_Q \sim 3 \times 10^7 \yr$ for two quasars.  However, they may be inconsistent with \citet{kirkman08}, who used a large sample of quasars with $\sim \Mpc$ separations to search the \ion{H}{1} forest for the transverse proximity effect.  They found no evidence for enhanced transmission in front of the quasars but \emph{decreased} transmission behind them.  They hypothesize that the increased gas density around the quasar hosts may cancel the expected increase in transmission in front of the quasars; in that case, the decreased transmission behind the foreground objects implies that the quasar light has not yet reached these regions, which in turn implies a short lifetime (or at least variability timescale) $\sim 10^6 \yr$.  Our constraints on variability on these very short timescales depend on the treatment of the data and so require more careful investigation.

There are, in addition, many simplifications in our model.  In addition to the inevitable measurement uncertainties in real experiments, we neglect the detailed structure of the forest and radiative transfer, which can induce small-scale features in the background.  In the finite lifetime case, the attenuation length of \ion{He}{2}-ionizing photons is an important parameter, but one that is very difficult to measure directly.  We have also assumed that optical surveys suffice to identify UV-bright quasars, but in reality there is substantial scatter in their far-UV properties \citep{telfer02, scott04}, which will degrade the correlation. 

Although we have focused on what the proximity effect can reveal about quasar lifetimes and beaming, there is more interesting physics to be gleaned.  For example, seeing any proximity effect at all implies that a quasar can not be ``flickering" more rapidly than the equilibration time of the gas ($\sim 3 \times 10^6 \yr$ at $\Gamma \sim 10^{-14} \secinv$, near the expected cosmic mean), or the mismatch from time delays wipes out the proximity effect.  Because this timescale is much longer than that for \ion{H}{1}, the resulting ``flickering" timescale is actually significant compared to our expectations for quasar lifetimes.  Thus the \ion{He}{2} forest can be used to measure long-term variability and perhaps even to constrain popular models like an exponentially decreasing quasar luminosity or the more complex light curves of \citet{hopkins05a}.  In either case, we might expect to find low-luminosity quasars (in their late stages of existence) correlated with surprisingly strong peaks in the hard ionizing background.  This may be very interesting in light of the rapid variability timescales suggested by \citet{kirkman08}.

Another possibility, in the context of finite quasar lifetimes, is to exploit the bare peaks by looking for a correlation between these and ``post-quasar" galaxies, perhaps with a recently concluded burst of star formation or evidence for recent strong mechanical feedback from the quasar. Alternatively, in a beaming model, one would instead search for correlations with galaxies showing signatures of obscured quasars, such as rapid ongoing star formation.

\acknowledgments

We thank C.~A. Faucher-Gigu{\`e}re and F. Davies for helpful discussions.  This research was partially supported by the NSF through grant AST-0829737 and by the David and Lucile Packard Foundation (SRF). 

\bibliographystyle{apj}
\bibliography{Ref_composite}

\end{document}